\documentclass{article}
\usepackage{spconf,amsmath,graphicx}
\usepackage[T1]{fontenc} 
\usepackage[utf8]{inputenc} 
\usepackage{graphicx}
\usepackage{balance}
\usepackage{color}
\usepackage[dvipsnames]{xcolor}
\usepackage{booktabs}
\usepackage{tabularx} 
\usepackage{multirow}
\usepackage{makecell}
\usepackage{amssymb,amsmath,mathtools}
\usepackage{amsthm}
\usepackage[ruled,vlined,linesnumbered,resetcount]{algorithm2e}
\usepackage{url}
\usepackage{siunitx}

\usepackage{harmony}

\usepackage[normalem]{ulem}
\useunder{\uline}{\ul}{}

\newcommand{\norm}[1]{\left\lVert#1\right\rVert}

\title{Towards Explaining Expressive Qualities in Piano Recordings: \\ Transfer of Explanatory Features via Acoustic Domain Adaptation}
\name{Shreyan Chowdhury$^1$\hspace{1cm}Gerhard Widmer$^{1,2}$
}
\address{$^1$Institute of Computational Perception and $^2$ LIT AI Lab, Johannes Kepler University Linz, Austria}
\begin{document}
%
\maketitle
\begin{abstract}
Emotion and expressivity in music have been topics of considerable interest in the field of music information retrieval. In recent years, \textit{mid-level perceptual features} have been suggested as means to explain computational predictions of musical emotion. We find that the diversity of musical styles and genres in the available dataset for learning these features is not sufficient for models to generalise well to specialised acoustic domains such as solo piano music. In this work, we show that by utilising unsupervised domain adaptation together with receptive-field regularised deep neural networks, it is possible to significantly improve generalisation to this domain.
Additionally, we demonstrate that our domain-adapted models can better predict
and explain expressive qualities in classical piano performances, as perceived and
described by human listeners.

\end{abstract}
\section{Introduction}

\textit{Domain mismatch} -- a discrepancy between the kind of data available for training
a classifier and the data on which it should then operate -- is an important
real-world problem, also in the field of acoustic recognition.
For instance, the DCASE 2019 and 2020 challenges had
dedicated tasks on \textit{Acoustic Scene Classification with
multiple/mismatched recording devices}.\footnote{e.g., http://dcase.community/challenge2019/task-acoustic-scene-\\classification-results-b}
The machine learning answer to this problem is research on effective methods
for \textit{transfer learning} and (supervised and unsupervised) 
\textit{domain adaptation}. 

The work presented in this paper is motivated by a particularly
difficult acoustic transfer problem involving a complex musical phenomenon.
In a large project,\footnote{https://www.jku.at/en/institute-of-computational-perception/research/\\projects/con-espressione}  we aim at studying the elusive concept of \textit{expressivity in
music} with computational and, specifically, machine learning methods.
One aspect of that is the art of \textit{expressive performance}, the subtle, continuous
shaping of musical parameters such as tempo, timing, dynamics, and articulation
by experienced musicians, while playing a piece,
in this way imbuing the piece with particular expressive and emotional qualities 
\cite{cancino2018computational}. 
The \textit{Con Espressione Game} was a large-scale data collection effort we
set up in order to obtain personal descriptions of perceived \textit{expressive qualities},
with the goal of studying human perception and characterisation of expressive aspects
in performances of the same pieces by different artists \cite{cancino2020characterization}. 

In analyzing this data, we are now interested in seeing whether these subjective characterisations of expressive qualities are
consistent and systematic enough for a machine to be able to predict them --
at least partially -- from the audio recordings.
Moreover, we aim at obtaining musical insights: we want
\textit{interpretable} models that point to specific musical qualities that might
underlie perceived expressive qualities.
A set of musical descriptors that seem particularly suited to this
was proposed in \cite{aljanaki2018}, where \textit{mid-level musical features}
were described that are intuitively understandable to the average musical
listener, and a corresponding human-annotated set of music
recordings was published (see next section).
In \cite{chowdhury2019towards}, we had shown how such mid-level features, predicted from
audio via trained classifiers, can be exploited to provide intuitive explanations
in the context of emotion and mood recognition in general (non-classical)
music.
This was extended to a two-level explanation scheme in \cite{haunschmid2019two}
which permitted to trace the mid-level explanations back to properties of the
acoustic signal.
There is reason to believe that some of these features may also hold
predictive and explanatory power for expressive aspects in piano performance.

This is where a severe \textit{mismatch problem} arises: 
there is no annotated ground truth data available for training mid-level feature
extractors in classical piano music, and obtaining such data
would be extremely cumbersome.
At the same time, recordings of solo piano music are very different,
musically and acoustically, from the kind of rock and pop music contained in
the available mid-level training dataset. It is thus likely that a classifier
trained on the latter will not generalise well to our
piano recordings.\footnote{Note that we cannot test this directly, as we
have no mid-level feature ground truth for the \textit{Con Espressione} performances. (We will use the few piano recordings in the midlevel dataset as our domain adapatation test set -- see Section \ref{sec:results}.)}

In this paper, we present several steps to bridge this domain mismatch through architecture choice and unsupervised domain adaptation techniques, and show that they are effective in generalising a model to solo piano recordings.
In a final step, we will try the adapted classifier on the
Con Espressione recordings, testing whether domain adaptation improves
the predictability of expressive qualities from mid-level features predicted
from audio, and identifying those features that seem to have specific predictive
and explanatory power.

\section{Mid-level Features and the \\ Con Espressione Data}

\subsection{The Mid-level Features Dataset}

Seven mid-level musical features were proposed in \cite{aljanaki2018}, viz. \textit{melodiousness, articulation, rhythmic complexity, rhythmic stability, dissonance, tonal stability}, and \textit{modality} (or ``minorness''). To approximate these perceptual features for a set of audio clips, the authors took a data-driven approach. Ratings from listeners were modelled into the final feature values that were made available as labels in the associated dataset (which we call the \textit{Mid-level Features Dataset}) along with the audio clips. The labels for each feature are in the form of continuous values between 1 and 10 (the learning task for our models will thus be a regression task.)
The exact questions asked to the listeners for rating each perceptual feature can be found in \cite{aljanaki2018}. The audio clips chosen for the dataset come from different sources such as \url{jamendo.com}, \url{magnatune.com}, and the Soundtracks dataset \cite{eerola2011}. There are a total of 5,000 clips of 15 seconds each in the dataset.

\subsection{The \textit{Con Espressione Game} Dataset}

In the \textit{Con Espressione Game}, participants listened to extracts from recordings of selected solo piano pieces (by composers such as Bach, Mozart, Beethoven, Schumann, Liszt, Brahms) by a variety of different famous pianists (for details, see \cite{cancino2020characterization}) and were asked to describe, in free-text format, the \textit{expressive character} of each performance. Typical characterisations that came up were adjectives like ``cold'', ``playful'', ``dynamic'', ``passionate'', ``gentle'', ``romantic'',
``mechanical'', ``delicate'', etc. From these textual descriptors, the authors obtained, by statistical analysis of the occurrence matrix of the descriptors, four underlying continuous expressive dimensions along which the performances can be placed. These are the (numeric) target dimensions that we wish to predict via the route of mid-level features predicted from the audio recordings.

The \textit{central challenge} in this is that the Mid-level Features Dataset \cite{aljanaki2018}, consisting mainly of 
pop, rock, hip-hop, jazz, electronic, and film soundtrack music, is vastly different, in sound and musical style, from the music of the Con Espressione dataset. This results in what is known as a \textit{covariate shift} \cite{ben2010theory} between the training and the testing data. 

In the following section, we describe a deliberate choice of training architecture that results in better generalisability of the trained models, and then present a two-step method to further adapt the model to our domain of choice.

\section{Mid-level Feature Learning \\ via Domain Adaptation (DA)}
In the following sections, \textit{target domain} refers to solo piano performance audio and \textit{source domain} refers to all other musical audio (non-piano audio clips in the Mid-level Features Dataset).

\subsection{Receptive Field Regularized ResNet}\label{ss:rfresnet}

As a first step towards improving out-of-domain generalization of mid-level feature prediction, we switch from the VGG-ish network of \cite{chowdhury2019towards} to the Receptive-Field Regularized ResNets (RF-ResNet) originally introduced in \cite{koutini2019receptive} for acoustic scene classification and later shown to work well for music information retrieval tasks as well \cite{koutini2019emotion}. The rationale behind this is that the smaller receptive field of the RF-ResNet prevents overfitting, particularly when the training data is limited in quantity. The architecture differs from a regular ResNet \cite{he2016deep} by reducing the kernel sizes of several convolutional layers and removing some max pooling layers. Our RF-ResNet consists of three stages with three residual blocks in the first stage and one residual block each in the second and third stages. The last stage consists of only 1-by-1 convolutional layers. There are two max pooling layers in the first stage between the convolutional blocks, and one average pooling layer after the third stage before going into a final 1-by-1 convolutional feed forward layer. The output is a seven-dimensional vector where the elements correspond to the predictions of each of the seven mid-level features.

\subsection{Unsupervised DA through Backpropagation}\label{ss:uda}

We adopt the \textit{reverse-gradient} method introduced in \cite{ganin2015unsupervised}, which achieves domain invariance by adversarially training a domain discriminator attached to the network being adapted, using a gradient reversal layer. The procedure requires a large unlabelled dataset of the target domain in addition to the labelled source data. The discriminator tries to learn discriminative features of the two domains but due to the gradient reversal layer between it and the feature extracting part of the network, the model learns to extract domain-invariant features from the inputs. 

This adaptation procedure is applied to the RF-ResNet described above. Since our target domain of interest solo piano performance music, we use audio from the MAESTRO dataset \cite{hawthorne2018enabling} as our unlabelled data source. It contains more than 200 hours of recorded piano performances. During training, each batch that the model sees contains an equal number of labelled source data points and unlabelled target data points. The regressor/classifier head of the model tries to predict the source labels while the discriminator head predicts the domain for each data point in the batch. The combined loss of the two heads is then backpropagated while reversing the gradient after the discriminator during the backward pass.

\subsection{Teacher-Student Training Scheme}

\begin{figure}[t]
\centerline{\includegraphics[width=0.95\columnwidth]{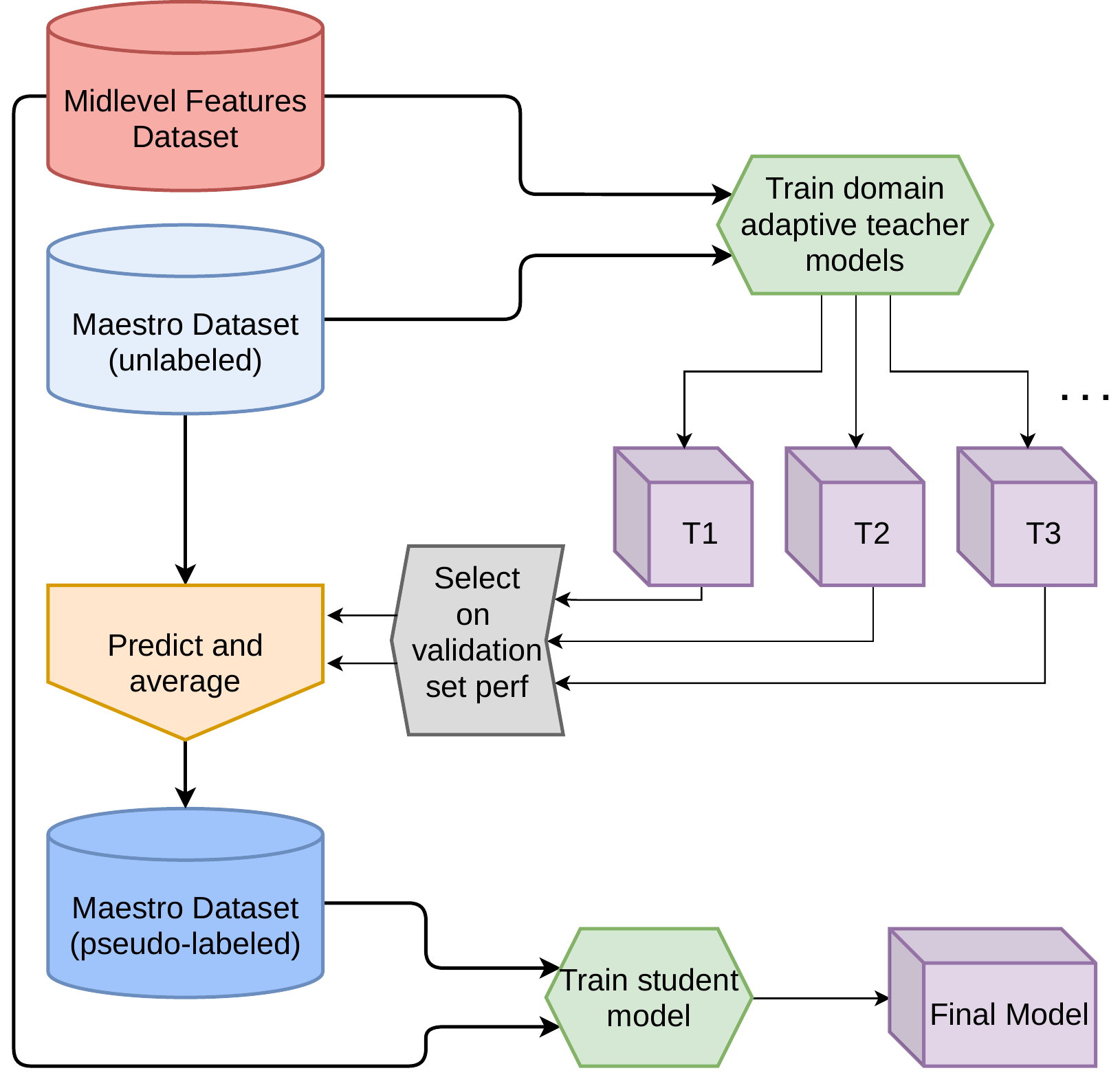}}
\caption{Teacher-Student training scheme for unsupervised domain adaptation.}
\label{fig:ts}
\end{figure}

As a final step, we refine our domain adaptation using a teacher-student training scheme
tailored to our scenario (see Fig.\ref{fig:ts}). We train multiple domain-adaptive models using the unsupervised DA method of Section \ref{ss:uda} and use these as teacher models that are eventually used to assign pseudo-labels to our unlabelled MAESTRO dataset. Before the pseudo-labelling step, we select the best performing teacher models with the validation set. Even though the validation set contains data from the source domain, this step ensures that models with relatively lower variance are used as teachers. This helps filter out the particularly poorly adapted models from the previous step, which may occur due to the inherently less stable nature of adversarial training methods \cite{che2016mode}.

After selecting a number of teacher models (in our experiments, we used four), we label a randomly selected subset of our unlabelled dataset using predictions aggregated by taking the average. This pseudo-labelled dataset is combined with the original labelled source dataset to train the student model. We observed that the performance on the test set increased until the pseudo-labelled dataset was about 10\% of the labelled source dataset in size, after which it saturated.

The teacher-student scheme allows the collective ``knowledge'' of an ensemble of adapted networks to be distilled into a single student network. The idea of knowledge distillation, which was originally introduced for model compression in \cite{hinton2015distilling}, has been used for domain adaptation in a supervised setting previously in \cite{asami2017domain}. The distillation process functions as a regularizer resulting in a student model with better generalisability than any of the individual teacher models alone. Additionally, it can be thought of as a stabilisation step helping to filter out the adversarially adapted models that result from non-optimal convergence.

\section{Experimental Results}
\label{sec:results}

Since we have no ground truth labels for our real data of interest (the classical piano music) to evaluate the domain adaptation experiments\footnote{\url{https://gitlab.cp.jku.at/shreyan/midlevel_da}}, we
created a (``piano''/target) test set manually by selecting clips from the Mid-level Features Dataset containing only solo piano. This resulted in a set of 79 piano clips from the total of 5000. The other 4921 clips (``non-piano''/source) were split into training (90\%), validation (2\%) and test (8\%) sets such that the artists in these sets are mutually exclusive (following \cite{aljanaki2018}). The validation set is used to tune hyperparameters and for early stopping.

The inputs to all our models were log-filtered spectrograms (149 bands) of 15-second audio clips sampled at 22.05 kHz with a window size of 2048 samples and a hop length of 704 samples, resulting in 149$\times$469-sized tensors. For training, we use the Adam optimizer with a multi-step learning rate scheduler. 
In the unsupervised DA step, the recordings from the MAESTRO dataset are split into 15-second segments and a random subset of size equal to the mid-level training set is sampled on each run. During the pseudo-labelling stage, a random subset of 500 segments is sampled. 

\begin{figure}[b]
\centerline{\includegraphics[width=\columnwidth]{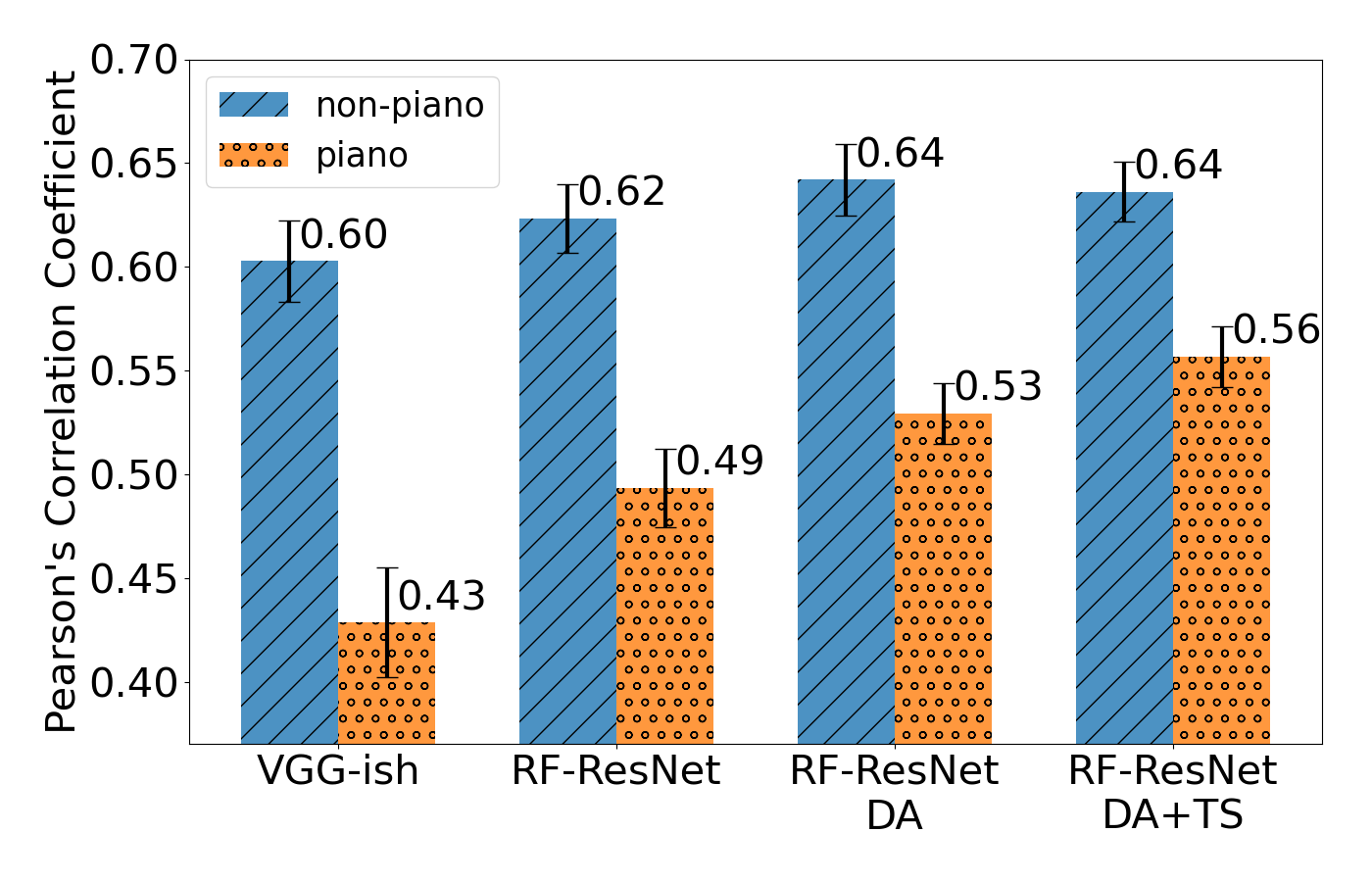}}
\vspace{-0.5cm}
\caption{Performance of mid-level feature models on non-piano and piano test sets.
}
\label{fig:summary}
\end{figure}

We observe (Fig. \ref{fig:summary}) that each of the steps mentioned in the previous section results in an improvement in the performance on the ``piano'' test set without compromising the performance on the ``non-piano'' one. In fact, we see a slight improvement in the non-piano metric upon introducing DA. This could be due to the presence of some data points similar to the target domain -- for instance excerpts from piano concertos, which are not included in the ``piano'' test set.

\begin{figure}[t]
\centerline{\includegraphics[width=\columnwidth]{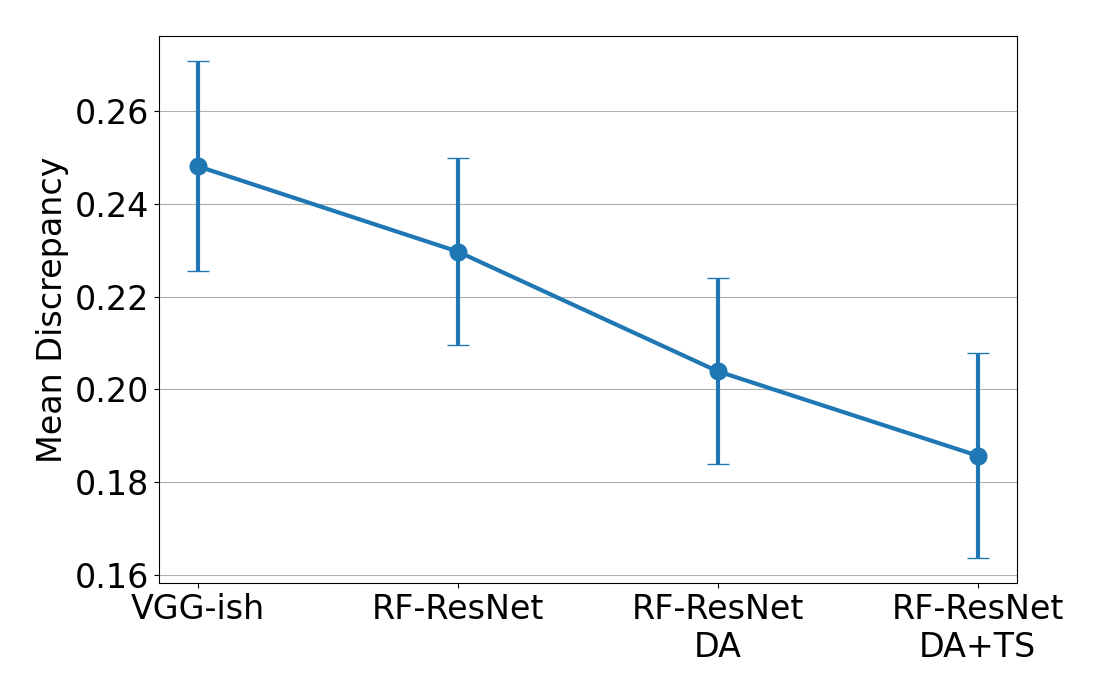}}
\vspace{-0.3cm}
\caption{Mean discrepancy between piano and non-piano sets.}
\vspace{-1mm}
\label{fig:mmd}
\end{figure}

To investigate our results further, we look at the discrepancy between the source and target domains in the representation space, since it is known that the performance of a model on the target domain is bounded by this discrepancy \cite{ben2010theory}. We use the method given in \cite{sun2019unsupervised} to compute the empirical distributional discrepancy between domains for a trained model $\phi$, which is given as $D(S',T';\phi)$ in Eq. \ref{eq:mmd}:
\vspace{-0.03cm}
\begin{equation}
    D(S',T';\phi) = \norm{\frac{1}{m} \sum_{x \in S'}\phi(x) - \frac{1}{n}\sum_{x \in T'}\phi(x)}_2
\label{eq:mmd}
\end{equation}

where 
$S'$ is a population sample of size $m$ from the source domain and $T'$ is a population sample of size $n$ from the target domain.
We observe that the discrepancy decreases for each step (Fig.\ref{fig:mmd}), justifying
our three-step approach and explaining the improvement in performance.

\section{Putting it to the Test}

As a final step, we now return to our real target domain of interest and briefly investigate whether our domain-adapted models can indeed predict better mid-level features for modelling the expressive descriptor embeddings of the Con Espressione dataset. We do this by predicting the average mid-level features (over time) for each performance using our models and training a simple linear regression model on these features to fit the four embedding dimensions. Even though this is a very abstract task, for a variety of reasons -- the noisy and varied nature of the human descriptions; the weak nature of the numeric dimensions gained from these; the complex and subjective nature of expressive music performance -- it can be seen that the features predicted using domain-adapted models give comparatively better R$^2$-scores for all four dimensions.

\begin{table}[t]
\centering
\begin{small}
\begin{tabular}{rllll}
\toprule
& Dim 1 & Dim 2 & Dim 3 & Dim 4 \\
\midrule
VGG-ish & 0.35 & 0.10 & 0.22 & 0.32 \\
RF-ResNet & 0.36 & 0.07 & 0.28 & 0.33 \\
RF-ResNet DA & \textbf{0.40} & 0.09 & \textbf{0.29} & 0.32 \\
RF-ResNet DA+TS & 0.35 & \textbf{0.15} & \textbf{0.29} & \textbf{0.34} \\
\bottomrule
\end{tabular}

\end{small}
\caption{Coefficient of determination (R$^2$-score) of description embedding dimensions of the Con Espressione game using a linear regressor trained on predicted mid-level features.}
\label{tab:expressive_dimensions}
\end{table}

\begin{table}[h]
\centering
\begin{small}
\begin{tabular}{lc|lc}
\toprule

\multicolumn{2}{c}{RF-ResNet} &
\multicolumn{2}{c}{RF-ResNet DA+TS} \\
Feature & $r$ & Feature & $r$ \\
\hline
articulation & 0.47 & melodiousness & $-$ 0.39  \\

rhythmic complexity & 0.41 & articulation & 0.46  \\

 &  & rhythmic complexity & 0.41  \\

 &  & dissonance & 0.40  \\
 
\bottomrule
\end{tabular}

\end{small}
\caption{Pearson's correlation ($r$) for mid-level features with the first description embedding dimension, with (right) and without (left) domain adaptation. Features with $\text{p}<0.05$ and $|r|>0.20$ are selected. This dimension has positive loadings for words like ``hectic'', ``irregular'', and negative loadings for words like ``sad'', ``gentle'', ``tender''.}
\label{tab:dim1}
\end{table}

Taking a closer look at Dimension 1 -- the one that came out most clearly in
the statistical analysis of the user responses and was characterized by descriptions
like ``hectic'' and ``agitated'' (as opposed to, e.g., ``calm'' and ``tender''; see \cite{cancino2020characterization}) -- 
and looking at the individual mid-level features (see Table \ref{tab:dim1}), we find that,
first of all, the predicted features that show a strong correlation with this
dimension do indeed make sense: one would expect articulated ways of
playing (e.g., with strong \textit{staccato}) and rhythmically complex or uneven
playing to be associated with an impression of musical agitation. 
What is more, after domain adaptation, the set of explanatory features grows,
now also including perceived dissonance as a positive, and perceived melodiousness
of playing as a negative factor -- which again makes musical sense and testifies
to the potential of domain adaptation for transferring explanatory acoustic and musical features.
\vspace{-3mm}

\section{Conclusion}
In this paper, we presented a three-step approach to adapt mid-level models for recordings of solo piano performances. We significantly improved the performance of these models on piano audio by using a receptive field regularised network and performing unsupervised domain adaptation via a teacher-student training scheme. We also demonstrated improved prediction of meaningful perceptual features corresponding to expressive dimensions. We conclude that this route of domain adaptation shows potential for a more general task of adapting models to specific genres or musical styles.

\section{Acknowledgment}
%
This work is supported by the European Research Council (ERC) under the
EU’s Horizon 2020 research \& innovation programme under grant
agreement No. 670035 (“Con Espressione”), and the Federal State of Upper Austria
(LIT AI Lab).

\vfill\pagebreak

\bibliographystyle{IEEEbib}
\balance
\bibliography{strings,refs}

\end{document}